# Evaluating Temporal Patterns in Applied Infant Affect Recognition

Allen Chang*
Computer Science Department
University of Southern California
Los Angeles, USA
changall@usc.edu

Lauren Klein*
Computer Science Department
University of Southern California
Los Angeles, USA
kleinl@usc.edu

Marcelo R. Rosales
Division of Biokinesiology and Physical Therapy
University of Southern California
Los Angeles, USA
mrrosale@usc.edu

Weiyang Deng
Division of Biokinesiology and Physical Therapy
University of Southern California
Los Angeles, USA
weiyangd@usc.edu

Beth A. Smith
Infant Neuromotor Control Lab
Children's Hospital Los Angeles
Los Angeles, USA
bsmith@chla.usc.edu

Maja J. Matarić
Computer Science Department
University of Southern California
Los Angeles, USA
mataric@usc.edu

*Abstract*—Agents must monitor their partners' affective states continuously in order to understand and engage in social interactions. However, methods for evaluating affect recognition do not account for changes in classification performance that may occur during occlusions or transitions between affective states. This paper addresses temporal patterns in affect classification performance in the context of an infant-robot interaction, where infants' affective states contribute to their ability to participate in a therapeutic leg movement activity. To support robustness to facial occlusions in video recordings, we trained infant affect recognition classifiers using both facial and body features. Next, we conducted an in-depth analysis of our best-performing models to evaluate how performance changed over time as the models encountered missing data and changing infant affect. During time windows when features were extracted with high confidence, a unimodal model trained on facial features achieved the same optimal performance as multimodal models trained on both facial and body features. However, multimodal models outperformed unimodal models when evaluated on the entire dataset. Additionally, model performance was weakest when predicting an affective state transition and improved after multiple predictions of the same affective state. These findings emphasize the benefits of incorporating body features in continuous affect recognition for infants. Our work highlights the importance of evaluating variability in model performance both over time and in the presence of missing data when applying affect recognition to social interactions.

*Index Terms*—affect recognition, multimodal interaction, time-continuous prediction

## I. INTRODUCTION

Across applications of affect recognition for infants, predictive models may be subject to different requirements with respect to time. Researchers in infant development study transitions in infants' and mothers' affect to evaluate relationships between infant-mother synchrony and developmental outcomes [1]. In this context, models must recognize the time of each change in affect to support analysis of interaction dynamics. Research exploring Socially Assistive Robotics (SAR) for infants [2] has identified affect as a key indicator of an infant's ability to engage with the robot. To tailor its actions to an infant's affective state, a SAR system must maintain a continuous prediction of affect even when data are missing. Evaluating affect recognition approaches in the context of missing data and affective state transitions is necessary for assessing their ability to support real-world applications.

Past work [3], [4] has analyzed infant cries to evaluate affect from segments of audio data. Additionally, video-based methods [5], [6] have extracted facial landmarks to classify infant affect from individual frames. As facial features are sometimes occluded, classification has been performed on frames without missing data. While these approaches support the potential of affect recognition for infants, performance metrics were not reported during times when data were missing or during affect transitions. As occlusions and changes in affect are often unavoidable, further work is needed to evaluate how automated methods perform when making predictions continuously in the context of infant social interactions.

Using a labeled video dataset collected during infant-robot interaction (IRI) studies conducted by our group [7]–[9], this paper addresses challenges to continuous infant affect recognition in two ways. First, we explore body features as an input to evaluate how using multiple modalities supports continuous affect prediction in the presence of missing data from either modality. Next, we explore trends in model performance over time. To analyze how the infants' past behaviors informed their current affective state, we repeated model training for a range of input window lengths. Using the highest performing unimodal and multimodal models, we analyzed performance with respect to time since actual and predicted transitions in infant affect. Specifically, we performed binary classification to recognize alert versus fussy infant states, as fussiness impacted infants' abilities to engage in the interactions.

This work is supported by the National Science Foundation under grant NSF CBET-1706964.
*equal contribution

Our results supported the use of body features in infant affect recognition. Unimodal models trained on body features achieved performance above random chance, with an average AUC of 0.70. While the best-performing face models (0.86 AUC evaluated on data with high feature extraction confidence, 0.67 AUC evaluated on all data) and multimodal models (0.86, 0.73) had equal AUC when evaluated on data with high feature extraction confidence, both multimodal models outperformed both unimodal models when evaluated on the entire dataset. For unimodal and multimodal models, we observed changes in the mean and variance of model accuracy with respect to the time since an actual or predicted state transition. Classification performance was weaker when predicting that an infant had changed affective states and improved after making consecutive predictions of the same state for several seconds. In the context of SAR, this result highlights a trade-off between waiting to achieve a high prediction certainty and missing an opportunity to intervene when infants become fussy. This work underscores the benefits of body features in automated infant affect recognition, as well as the importance of considering temporal trends in model performance. The main contributions of this work are as follows:

1) We demonstrate body features as a viable modality for predicting infant affect.
2) We analyze temporal factors that impact model performance, including input window length, time since the last transition between affective states, and time since the last transition between classifier predictions.

## II. RELATED WORK

### A. Automated Detection of Infant Emotion Expression

Infant affect is commonly observed during social interactions to study phenomena such as pre-term birth, maternal depression, and later diagnosis of autism spectrum disorder, among others [1], [10]. Manual coding is labor intensive, posing a scalability challenge; therefore, researchers have begun exploring automated approaches for recognizing infant affect. Cohen and Lavner [4] used voice activity detection and k-nearest neighbors to detect infant cries. Model performance was evaluated based on the fraction of 1-second and 10-second segments that were accurately classified. A review by Saraswathy et al. [3] describes methods for automatic analysis of pre-detected infant crying sounds used to scan for signs of adverse health outcomes. Lysenko et al. [5] and Messinger et al. [6] used facial landmarks to predict expressions of infant affect from individual video frames. Models evaluated on pre-defined windows of time have identified the potential of affect recognition in analyzing infant well-being. To extend this research to social interaction, where timing is essential, we evaluate the impact of temporal factors on model performance.

### B. Infants and Children in SAR

Recent work has leveraged advances in SAR to design therapeutic interventions that have shown potential to support infant development. As described in Section I, recent studies [7]–[9] demonstrated the use of SAR to encourage leg movement in infants. Data collection was stopped if an infant became too distressed, ending the opportunity to learn from the robot. Scassellati et al. [2] used a virtual avatar and robotic tutor to demonstrate sign language to deaf infants. They used thermal imaging to infer infants' emotional engagement in the activity. Shi et al. [11] trained personalized models to recognize child affect from video, audio, and game performance toward enabling SAR tutors to respond to children's affective cues. To intervene when human partners become emotionally unavailable to interact, robots must monitor affect continuously and recognize changes quickly.

### C. Analysis of Affect with Body Expression

Past research has found that the speed of body motions can describe different emotional states in adults. For example, anger may be indicated by rapid, jerky movements, whereas sadness is commonly manifested with slow, fluid movements [12]. Taking advantage of these findings, recent work in adult affect recognition has shown success when incorporating body expressions [13], [14]. Work in affective body expression recognition demonstrates that a combination of posture (e.g., leaning directions, head position) and movement (e.g., gesture speed) can discriminate between affective states.

Past work indicates that infants, too, demonstrate their affect through body postures, motions, and gestures. Sauter et al. [15] identified that gestures such as clapping and arm-waving occur simultaneously with positive affect in infants. Burt [16] described differences in movement directions across joy, interest, sadness, and anger. Lysenko et al. [5] identified correlations between wrist velocity and infant emotion expression. To the best of our knowledge, this work is the first to integrate body expressions into automated infant affect recognition.

## III. IRI DATASET

As described in Section I, the IRI dataset was collected during a series of infant-robot interaction studies. Each study involved a SAR interaction designed to encourage infants to make exploratory leg movements. These studies were part of a larger project to explore SAR as a method of observing and potentially supporting infant motor and cognitive development.

### A. Participants

The IRI dataset included 26 (19 female, 7 male) infants between 6 and 9 months of age with typical development from the greater Los Angeles area. Some infants participated in more than one study. Ethnicity was reported by each infant's parent, with 7 identified as Hispanic or Latino. Race was also reported by parents, with 2 identified as Asian, 1 as Black or African American, 13 as White, 9 as "other", and 1 parent declined to answer.

### B. Infant-Robot Interaction Setup

Infants were fastened in a seat across from the Nao humanoid robot as shown in Fig. 1. Separate cameras were used to film the infants' faces and bodies. During each interaction, the robot provided sensory feedback each time the infant moved their leg above an acceleration or angular velocity threshold. Specifically, the robot either kicked, kicked and flashed lights, or kicked and played an infant laughing sound to encourage the desired leg movement. Acceleration and angular velocity thresholds changed between studies to encourage different movement behaviors. More detailed information about the study setup can be found in prior work [7]–[9].

### C. Manual Affect Coding

Infant affect was manually coded using a five-point arousal scale consistent with previous infant research [17]. Annotators labeled each video frame as either alert, fussy, crying, drowsy, or sleeping. Annotators achieved over 85% label agreement. 84.3% of the data were labeled as alert, 13.4% as fussy, 2.3% as crying, and 0% as drowsy or sleeping. Given the limited data labeled as crying and similarities between fussy and crying, these labels were combined as fussy in our analysis.

### D. Exclusion Criteria

Data from 8 interactions were excluded from our experiments because the labels or face-view videos were not present in the IRI dataset. After exclusions, 230 minutes of video were included in our analysis.

## IV. METHODS

To understand how our modeling approach may support continuous affect recognition, it is necessary to evaluate both its overall performance and patterns in performance across time. In the context of the SAR interactions described in Section III-B, performance changes may influence a robot's level of certainty, and therefore its action selection policy. This section describes our feature extraction approach, affect classification models, and evaluation methods.

### A. Feature Extraction

We used OpenFace [18] to extract 68 facial landmarks and 17 action units (AUs) from the face-view videos, and OpenPose [19] to extract 25 body landmarks from the body-view videos. OpenFace and OpenPose were trained with adult data, so visual inspection was conducted to assess that the accuracy of model output was sufficient. When landmarks were not detected, their coordinates were 0. As infant leg movements directly controlled the SAR feedback, landmarks from the legs were discarded to mitigate differences across interaction difficulty. The method described by Klein et al. [20] was used to identify which body landmarks belonged to the infant when researchers or parents were present in the videos. Face-view videos had variable frame rates while body-view videos had frame rates of 29.97 frames per second. To mitigate differences in frame rates, mean landmark and AU values were aggregated across 0.25 second intervals. While OpenFace detects AUs using a sequence of frames, a similar dynamic representation was not available for body gestures, so we calculated the speed of each body landmark as in Yang et al. [21] to account for movement. A landmark's speed was calculated as the distance it traveled since the previous frame.

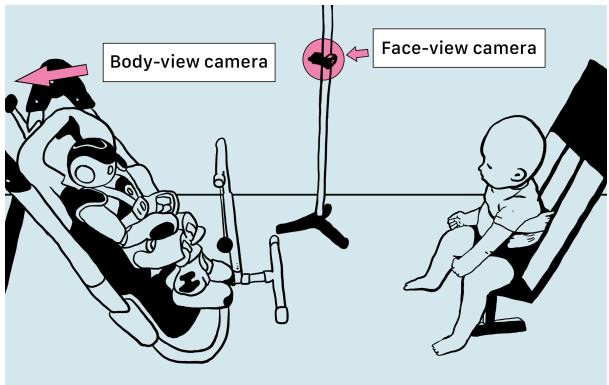

Fig. 1. Infant-robot interaction setup: infants were seated in a chair across from the seated Nao robot.

As occlusions were common, we used a 20% feature extraction confidence threshold to identify data unlikely to represent actual infant behavior while still retaining the majority of frames. 58.9% of OpenFace data and 78.6% of OpenPose data exceeded the confidence threshold. Visual inspection indicated that facial feature extraction with low confidence often occurred when infants leaned away from the camera or turned their heads sideways to look at their parents. This discrepancy in feature availability supports the integration of multiple data sources in recognizing infant affect.

As the SAR system feedback sometimes included infant laughing sounds, and parents or researchers occasionally spoke during the interaction, it was unclear whether extracted audio features would describe infant behavior. Therefore, audio data were not included in the scope of this work. Future work will explore methods for incorporating audio data in infant affect recognition while accounting for multiple sources of noise.

### B. Feature Preprocessing

Landmark positions were centered and scaled to address differences in infants' sizes and positions relative to the cameras. First, we identified a landmark that was approximately centered in the video and infrequently occluded and subtracted its coordinates from each landmark. The tip of the nose was selected as the centered landmark for the face and the neck was selected for the body. Coordinates were then scaled according to each infant's size in the video. Facial landmark coordinates were divided by the length of each infant's face, approximated as the distance between landmarks at the top and bottom of the head. Distances between body landmarks were scaled by the distance between the neck and the pelvis, an approximation for the length of the infant's torso. As in past work by Messinger et al. [6] and Lysenko et al. [5], we included the distances between each pair of landmarks as features.

### C. Temporal Feature Aggregation and Windowing

To incorporate temporal patterns in facial and body features, we aggregated data over windows of time as in Filntisis et al. [22]. We calculated the maximum, mean, and standard deviation of features over each window. This method leveraged temporal information without the use of recurrent neural networks, which are prone to overfitting on small datasets [23], and produced fewer parameters for training compared to flattening the feature input. However, aggregation removes local temporal patterns. To address this trade-off, two time windows were applied to capture both short term and long term trends, as in action recognition work by Yang et al. [21]. As real-time classifiers can only leverage past time steps, the windows were aligned and labeled at the end of each time window. Speed was calculated based on differences in landmark positions between frames, so the short window needed to be at least 2 frames. Each frame was 0.25 seconds, so 0.5 seconds was chosen as the short window length. As there does not exist a standard window length describing the amount of past infant behavior needed to inform affect recognition, we tested long window lengths starting at 1 second and increasing by a factor of 2 until a peak in performance was found. We report results across this range to analyze the impact of windowing on model performance.

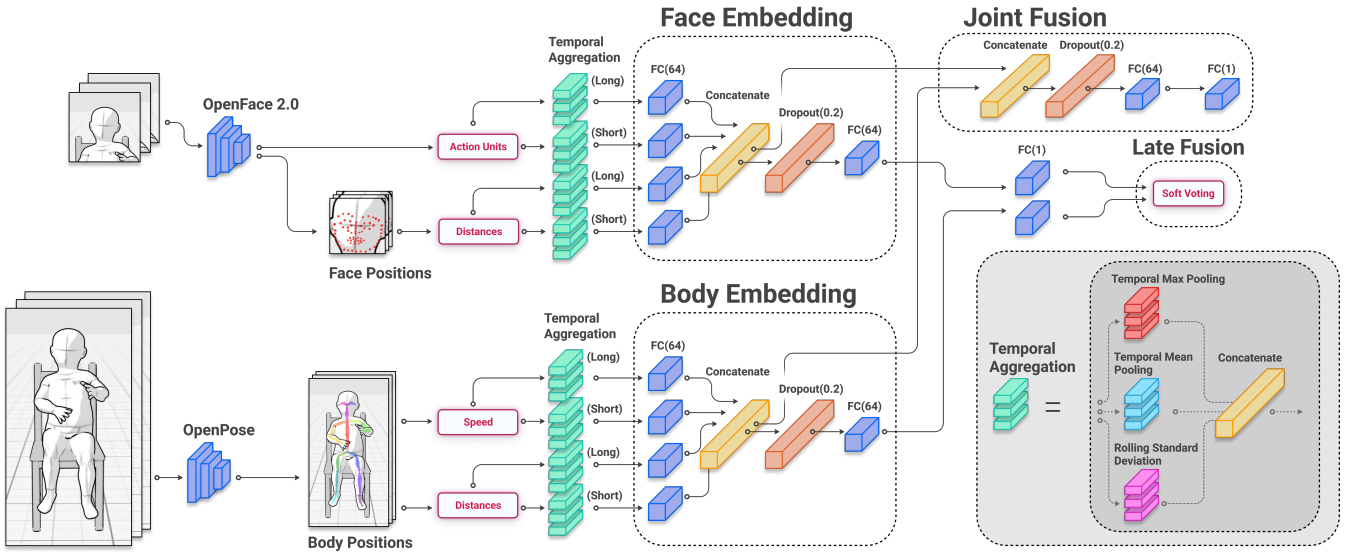

Fig. 2. Overview of the modeling framework. FC: Fully Connected (Dense) Layer.

Window lengths were varied independently for facial and body features to explore differences across modalities. Samples were considered successful if the feature extraction threshold was met for at least 90% of the long window to identify windows most likely to represent actual infant behavior without penalizing windows with a few noisy frames. In order to compare models on the same dataset, only samples that met the success criteria for the longest tested window size were used for model training. We label the dataset comprised of these samples as $\mathcal{D}_{confident}$ and the entire dataset as $\mathcal{D}_{total}$.

### D. Feature Selection

We categorized the temporally aggregated features into 4 distinct feature groups: facial landmark distances, facial AUs, body landmark distances, and body landmark speeds. Next, we applied Welch's unequal variances t-test to determine the aggregated features with the largest differences in means between labels. Models were fit using the 12 features of each group with the lowest p-values from the t-test to prevent overfitting while balancing feature representation across modalities. The feature selection process was performed independently for each training set during 5-fold cross validation.

### E. Affect Classification

*1) Model Architecture:* We trained neural networks (NN) for each window length described in Section IV-C. The NN architecture was based on work by Filntisis et al. [22] to classify children's affect using facial and body features. As illustrated in Fig. 2, the feature groups described in Section IV-D were transformed through individual fully connected (FC) layers with the ReLU activation function. The outputs of these layers were concatenated and fed through another FC layer. Applying separate FC layers allowed the NN to model relationships both within and between groups of features.

*2) Unimodal and Multimodal Classifiers:* First, unimodal models were trained only with facial features or body features, and are referred to in this paper as the face and body models. Next, we evaluated approaches for integrating both modalities. We leveraged a joint fusion model to account for interactions between facial and body features. The input channels were the same as those from our unimodal models and were fused at the concatenation layer before producing a final affect classification. To evaluate fusion at the decision level, we implemented a late fusion model using a soft voting approach that weighted the face and body models equally. While past work in multimodal affect recognition [24] has leveraged event detection to recognize modality-specific behaviors prior to fusion, that approach requires labeled behaviors in each modality, which were not available in the IRI dataset.

*3) Model Training and Evaluation:* Each model was initialized randomly without seeding and trained with binary cross-

entropy loss and the Adam optimizer. The ratio of fussy to alert labels was approximately 1:9, so we applied a bias to the loss function where the penalty was scaled by a factor of 9 for the fussy class compared to the alert class to help with training. Models were trained with 5 epochs to prevent overfitting. Results were evaluated using stratified 5-fold cross-validation, with each infant's data represented in only 1 fold. We report the mean AUC across the 5 folds for each model. AUC is a commonly used performance metric for binary classification and is invariant to prior class probabilities [25], making it appropriate for our imbalanced dataset. To visualize the information embedded by the model, we conducted principal component analysis with 2 dimensions on the output of the final embedding layer. The face and body principal components from 1 of the 5 test folds are shown Fig. 3.

### F. Trends in Model Performance Over Time

We evaluated the highest performing models on $\mathcal{D}_{total}$ to explore how performance changed 1) in the presence of missing data and 2) during actual or predicted changes in affect to evaluate the models in the context of a continuous interaction. This analysis was conducted separately for the face, body, joint fusion, and late fusion models.

*1) Accuracy Versus Time Since Affect Transition:* As infants switch between affective states, their behavior may evolve over time rather than abruptly. For example, infants may behave differently when they have been fussy for 1 second versus several seconds. These differences in behavior may, in turn, produce changes in model performance over time. Therefore, we investigated whether model performance varied as more time elapsed after a change in an infant's affective state. We abbreviate "time elapsed since an affective state transition" as TSAT. For each infant, we separated data instances by affective state label and then grouped instances that were within 0.25 seconds of each other in TSAT. Across all infants, we calculated the average accuracy of our models on samples from each TSAT group to explore how the models performed at various times since the most recent change in infant affect.

*2) Accuracy Versus Time Since Predicted Transition:* A SAR system that responds to infant affect must evaluate the certainty of its affect classifications to inform its action selection policies. For example, if model performance is different when predicting a change in affective state versus predicting that an infant has remained fussy for several seconds, this may influence the robot's actions. To address this possibility, we explored whether model performance differed based on the number of consecutive predictions of the same affective state. For each infant and model, we grouped samples by the predicted affective state and by the number of seconds since the model last changed its prediction, at 0.25-second intervals. We label "time since a predicted state transition" as TSPT. To evaluate whether each model's accuracy changed after predicting the same label for multiple seconds, we calculated average model accuracy across each infant and affective state for each TSPT group.

## V. RESULTS AND DISCUSSION

### A. Accuracy Versus Input Window Length

For models trained on $\mathcal{D}_{confident}$, results indicated a longer optimal long window for facial features compared to body features. The difference suggests that infants in the IRI dataset displayed their affect along different time scales for each modality. As illustrated in Fig. 4, mean AUC for the face and multimodal models increased until the long input window length for facial features reached 32 seconds. Meanwhile, the body and joint fusion models had higher mean AUC for shorter time windows. The longer optimal window for facial features may have provided important context to compare against observations from the short window. A longer window did not provide consistent benefits for the body modality, which may indicate that infants' body features from several seconds ago were less relevant to their current affective state. This trend may serve as an advantage for models trained on body features, since a shorter time window is less affected by occlusions that occurred several seconds ago.

For models evaluated on data with high feature extraction confidence, mean AUC was higher for the face models compared to body models, and was similar between face models and the highest performing multimodal models. These scores indicate that when both sources of data were readily available for the past 64 seconds, facial features were equally or more informative than body features or a combined feature set, which is consistent with the illustration in Fig. 3, showing less overlap in the first two principal components of the facial embeddings compared to the body embeddings. The difference in ability to discriminate between alert and fussy affect may be related to the differences in available features. While facial features included AUs, we did not similarly incorporate indicators of body gestures. However, body models outperformed random chance with a mean AUC of 0.70 across the tested window lengths.

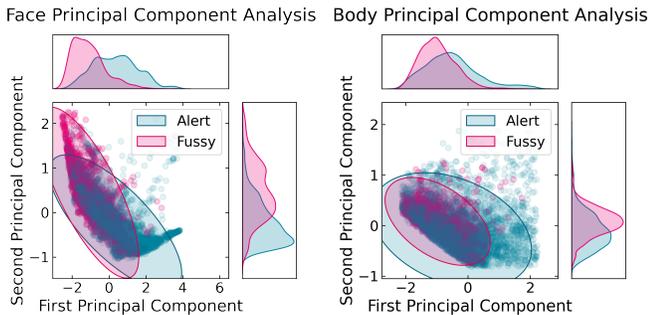

Fig. 3. Scatter plot and marginal density distributions of the first two principal components of face and body NN embeddings, for 1 test fold of infants. Ellipses contain data points within 3 standard deviations of the mean.

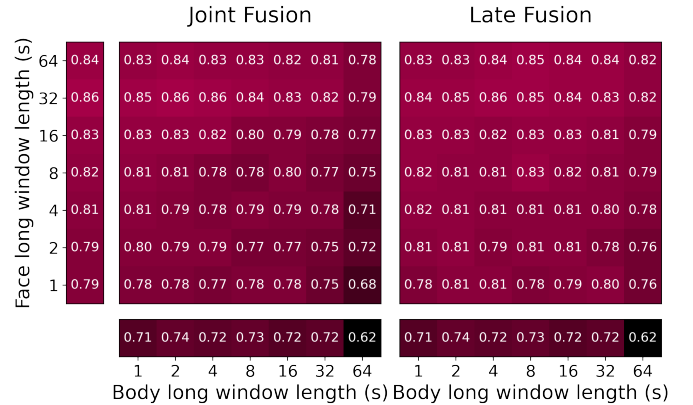

Fig. 4. Mean AUC across stratified cross-validation folds for each set of long window lengths for facial and body features.

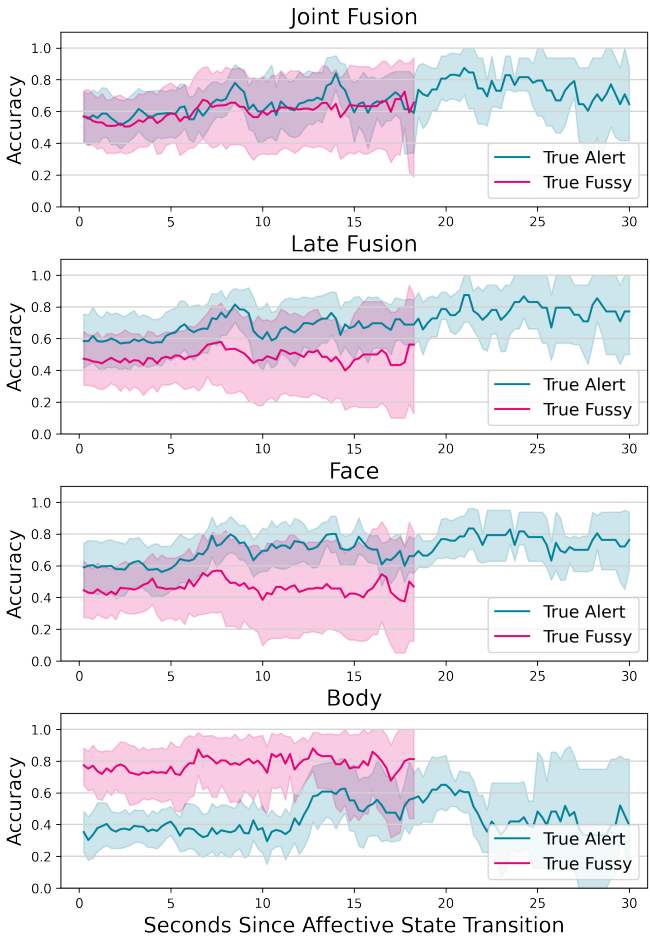
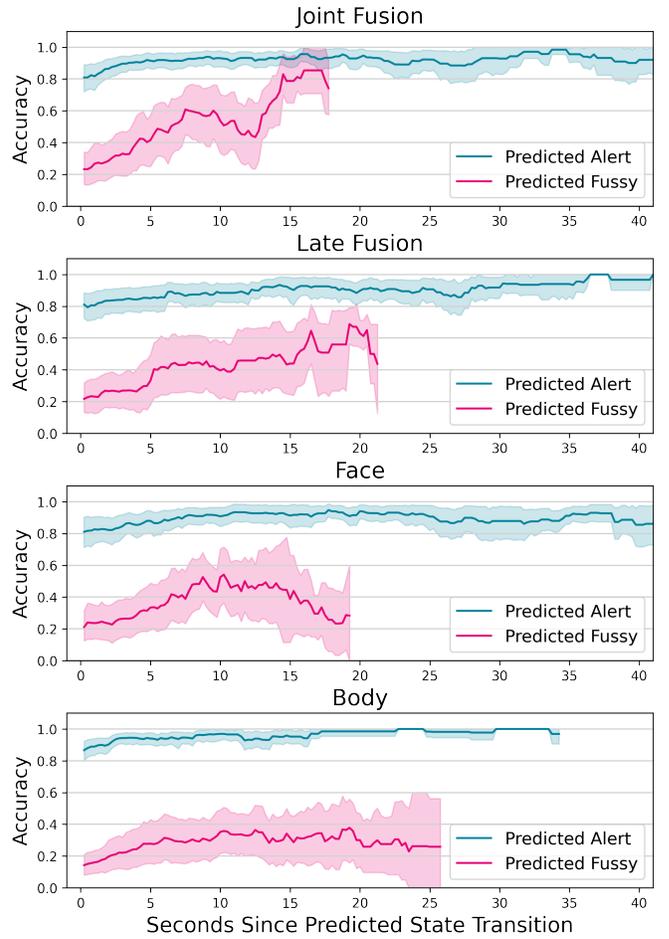

Fig. 5. Model accuracy (mean and 95% confidence interval) vs. seconds since infants transitioned into a given affect (TSAT). Only TSAT values with more than 3 samples are shown. The "True Alert" curve visualizes model accuracy for samples labeled as alert, while the "True Fussy" curve visualizes model accuracy for samples labeled as fussy.

Fig. 6. Model accuracy (mean and 95% confidence interval) vs. time since the last predicted affect transition (TSPT). Only TSPT values with more than 3 samples are shown. The "Predicted Alert" curve visualizes model accuracy for samples predicted as alert, while the "Predicted Fussy" curve visualizes model accuracy for samples predicted as fussy.

### B. Trends in Model Performance Over Time

We analyzed the performance of our highest scoring models, with long window lengths of 32 seconds for facial features and 2 seconds for body features, on $\mathcal{D}_{total}$. Mean AUC for the face, body, joint fusion and late fusion models were 0.67, 0.65, 0.73, and 0.69, respectively. In contrast to the models trained on $\mathcal{D}_{confident}$, both multimodal models outperformed both unimodal models when trained on $\mathcal{D}_{total}$. While data imputation may mitigate the impact of missing data, these results support the use of body features when estimating infant affect continuously during social interactions.

*1) Accuracy Versus Time Since Affect Transition:* For samples labeled as alert, we observed an increase in accuracy for face and multimodal models with respect to TSAT, as illustrated in Fig. 5. This suggests that, as the infants transitioned from fussy to alert, their changes in behavior occurred gradually over several seconds rather than immediately. An increase in accuracy was not observed for the body models. This highlights the impact of window length, as longer windows are more likely to incorporate data from a previous affective state. Additionally, we note that during times when infants were labeled as fussy, the 95% confidence interval expanded over time for models across all modalities but remained the same across TSAT when infants were labeled as alert. The trends in the confidence intervals support that infants behaved similarly after first becoming fussy, whereas when they remained fussy for long periods of time, the ways in which they expressed their frustration became more varied.

For the face and late fusion models, accuracy was higher when infants were labeled as alert. This may have reflected class imbalance, as the majority of data in the IRI dataset were labeled as alert. In contrast, the body models were more accurate when infants were labeled as fussy as compared to alert, and the joint fusion model performed similarly across affective labels. These differences may have reflected the impact of the class weights on the learned decision boundary.

*2) Accuracy Versus Time Since Predicted Transition:* Across all models, accuracy first increased with the number of consecutive predictions of the same affect, followed by a decrease after several seconds. These trends are illustrated in Fig. 6. This pattern may reflect gradual rather than sudden transitions as affect shifted between states. For samples labeled as alert, the decrease in mean accuracy did not occur until after 30 seconds for any model. In data labeled as fussy, this decrease occurred at different times across models. Unimodal models experienced a decrease in performance after predicting a label of fussy for approximately 10 consecutive seconds. Meanwhile, when multimodal models predicted a label of fussy, peak accuracy was reached after 15 seconds before experiencing a decrease in performance, and greater peak accuracy was achieved compared to unimodal models. Higher classification accuracy over time for multimodal models compared to unimodal models when evaluated on $D_{total}$ demonstrates the benefit of body features for continuous affect recognition in the presence of missing data.

Future work should address the relationship between the number of consecutive predictions and affect recognition performance. Whether considering infants or older populations, changes in model performance over time impact a system's ability to respond quickly to affective state transitions and to assess the duration of a given state. For example, a SAR system that intervenes when a human partner is upset should balance the rising certainty over time (as illustrated in Fig. 6) with the ramifications of waiting too long to provide support.

As in Section V-B1, we note that the classifiers demonstrated a lower mean and higher variance in performance when predicting a label of fussy rather than alert, and that class imbalance should be considered when applying affect recognition to tasks such as the automated evaluation of infant-mother interactions [1] or the design of action selection policies during SAR interactions. In SAR, evaluating the trade-off between overall model accuracy and recall of negative affective states is essential when forming an action selection policy. For example, a robot that over-predicts a label of fussy and errs toward unnecessary soothing behaviors may support more successful interactions compared to a robot that errs toward not intervening when infants are fussy.

## VI. CONCLUSION

This paper presents an analysis of temporal trends in unimodal and multimodal infant affect recognition performance using facial and body features. In addition to demonstrating body features as a viable modality for predicting infant affect, we explored how changes in the length of time included in input windows and the duration of an actual or predicted affective state are associated with classification performance. Longer optimal window lengths of facial features compared to body features suggest that infants in the IRI dataset displayed their affect along different time scales across modalities. Multimodal models outperformed unimodal models when evaluated on the entire dataset $D_{total}$ but not when evaluated on $\mathcal{D}_{confident}$. The mean and variance in classification accuracy changed with the duration of actual and predicted affective states. These results highlight the importance of testing models on continuous interaction data and assessing patterns in performance over time when applying affect recognition to real-world social interactions. Future work should evaluate how these results generalize across infant datasets and additional modalities. Taken together, the results of this work inform guidelines for the application of automated infant affect recognition in the context of social interaction.

### ETHICAL IMPACT STATEMENT

This research was approved by the USC Institutional Review Board under protocol #HS-14-00911.

Since our approach used de-identified features, our models can be applied without storing identifiable data. To support transparency and accountability, a datasheet following the standards described by Gebru et al. [26] is available at https://github.com/interaction-lab/Infant-Robot-Interaction.

Since half of the IRI dataset's participants are White, the training data may have biased our classifiers to better recognize affect for White infants than those of other racial groups. Future work should train models on data representing a larger and more inclusive population prior to application.

### CITATION DIVERSITY STATEMENT

Recent work in several fields of science has identified a bias in citation practices such that papers from women and other minority scholars are undercited relative to the number of papers in the field [27]–[31]. We recognize this bias and report estimated gender and ethnic breakdowns below. Excluding self-citations, our references contain 26% woman(first)/woman(last), 7% man/woman, 19% woman/man, 41% man/man, 4% man/non-binary, and 4% unknown. Our references contain 11% author of color (first)/author of color(last), 4% white author/ author of color, 4% author of color/white author, 74% white author/white author, and 8% unknown. While we used the resources available to us to determine author gender and race (such as personal websites, university faculty biographical statements, or online APIs suggested by [32]), there is a possibility of mislabeling, as we did not contact authors directly for confirmation. We look forward to future work that could help us to better understand how to support equitable practices in science.

### ACKNOWLEDGMENT

This work was supported by the National Science Foundation (NSF CBET-1706964).